\documentclass{article}
\usepackage{amsmath}
\usepackage[dvips]{graphicx}
\usepackage{amssymb}

\title{Overcritical $\mathcal{PT}$-symmetric square well potential in the
Dirac equation}

\author{Francesco Cannata $(^{a,b})$ and
Alberto Ventura $(^{b,c})$ \\
$(^a)$ Dipartimento di Fisica dell' Universit\`{a} di Bologna \\
$(^b)$ Istituto Nazionale di Fisica Nucleare, Sezione di Bologna \\
$(^c)$ Ente per le Nuove Tecnologie, l'Energia e l'Ambiente, Bologna}
\begin{document}
\maketitle
\begin{abstract}
We study scattering properties of a $\mathcal{PT}$-symmetric square well
potential with real depth larger than the threshold of particle-antiparticle
pair production as the time component of a vector potential in the Dirac
equation. Spontaneous pair production inside the well becomes tiny beyond
the strength at which discrete bound states with real energies disappear,
consistently with a spontaneous breakdown of $\mathcal{PT}$ symmetry.
\end{abstract}

\section{Introduction}

The study of $\mathcal{PT}$-symmetric potentials originated in the seminal
papers\cite{BB98},\cite{BBM99} by Bender and coworkers, dedicated to the
analysis of spectra of non-Hermitian Hamiltonians of anharmonic oscillators,
which turn out to be entirely real on condition that exact $\mathcal{PT}$
symmetry holds, \textit{i.e.}, the Hamiltonian, $H$, commutes with the $%
\mathcal{PT}$ operator and all eigenfunctions of $H$ are also eigenstates of 
$\mathcal{PT}$. In case of exact $\mathcal{PT}$ symmetry, it is possible to
formulate an equivalent quantum mechanical description by defining a new
metric operator in the representation space, a concept already established%
\cite{SGH92} before the study of the properties of $\mathcal{PT}$-symmetric
potentials and generalized later to the definition of pseudo-Hermitian
quantum mechanics\cite{Mo02}. If the condition that the eigenfunctions of
the $\mathcal{PT}$ invariant Hamiltonian are also eigenfunctions of $%
\mathcal{PT}$ is relaxed, the $\mathcal{PT}$ symmetry is spontaneously
broken and complex eigenvalues appear in the spectrum of $H$. Since then
many authors have examined $\ $bound state problems with $\mathcal{PT}$%
-symmetric potentials in non relativistic quantum mechanics, while
relatively few studies are dedicated to relativistic models, or even to $%
\mathcal{PT}$-symmetric quantum field theories\cite{BJR05, B07}. The present
work intends to be a contribution to the study of the $\mathcal{PT}$ \
invariant Dirac equation, with emphasis on scattering aspects, within the
framework of standard relativistic quantum mechanics. $\mathcal{PT}$%
-symmetric potentials will be treated phenomenologically as effective
potentials, without attempting to formulate an equivalent Hermitian theory 
\cite{B07}.

Solvable models in the Dirac equation are most easily constructed in (1+1)
space-time dimensions and the related $\mathcal{PT}$-symmetric potentials
may have different behaviour under Lorentz transformations, and are
classified as vectors, pseudovectors, scalars, or pseudoscalars. In the
cases of scalar and pseudoscalar potentials, affecting particles and
antiparticles in the same way, the relevant hidden symmetry of the Dirac
equation can be classified as pseudosupersymmetry\cite{SR05}. The analysis
of vector potentials has been focused mainly on bound states, like in the
generalized Hult\'{e}n potential of Ref.\cite{ES05}, \ or in the logarithmic
derivative of a suitable position-dependent effective mass of Ref.\cite{JD06}%
, \ while scattering aspects have not been examined in detail. Among the
latter, an interesting peculiarity of strong potentials, with $\left\vert
V\right\vert >2m$, the threshold for production of a particle-antiparticle
pair, with $m$ the particle mass, is the possibility that bound states merge
with the negative-energy continuum of scattering states, thus appearing as
transmission resonances at negative energies inside the potential well,
which becomes overcritical with respect to spontaneous creation of
particle-antiparticle pairs.

Such overcritical vector potentials are well studied in the Hermitian case
in (1+1) dimensions, from the simplest example of the square well\cite{Gr97}%
, to the cusp potential\cite{DR95},\cite{VG03}, or the Woods-Saxon potential%
\cite{Ke02}. In the present work, we study scattering aspects of
overcritical $\mathcal{PT}$ -symmetric potentials, in particular how
overcriticality is interrelated with spontaneous breakdown of $\mathcal{PT}$
symmetry. It is worthwhile to remark from the very beginning that our
framework is standard quantum mechanics with complex potentials,
phenomenologically treated as effective potentials. For the sake of
simplicity, the analysis is carried out in detail for the $\mathcal{PT}$%
-symmetric square well, taken as the time component of a vector potential.

In this short note, we shall not enter into discussion of construction of
positive definite norms via a linear operator, $\mathcal{C}$, commuting with
the $\mathcal{PT}$ operator and the Hamiltonian, $H$, introduced by Bender
and collaborators in nonrelativistic quantum mechanics\cite{BMW03},\cite%
{BBJ04},\cite{BT06} and later extended to quantum field theory, in
particular $\mathcal{PT}$ -symmetric quantum electrodynamics, with imaginary
electric charge and axial vector potential\cite{BCMS05}.

We already know from previous work on bound states in the 
Schr\"{o}dinger equation that the $\mathcal{PT}$ symmetry of the square well
is spontaneously broken~\cite{Zn01},~\cite{ZL01}
and that the perturbative derivation of the
$\mathcal{C}$ operator below the critical value of the imaginary part
of the potential is by no means trivial~\cite{BT06}. As far as scattering
states are concerned, however, we have pointed out in our previous work on non
relativistic scattering~\cite{CDV07}, that, even in the case of spontaneous
breakdown of $\mathcal{PT}$ symmetry, transmission and reflection coefficients
for progressive waves, travelling from left to right on the real axis, and for
regressive waves, travelling from right to left, have definite
non trivial relations, which are not
valid for a general complex potential. 
These relations are connected with the fact that the 
imaginary part of a $\mathcal{PT}$-symmetric potential is an odd function
of the space coordinate $x$, so that its integral over the $x$ axis vanishes.

The present work is thus intended as a phenomenological investigation 
of relativistic features, such as overcriticality, in presence of
a spontaneously broken  $\mathcal{PT}$ symmetry, 
which would hardly emerge
for different reasons, either in the case of a
general complex potential, or in the case of a potential with exact
asymptotic  $\mathcal{PT}$ symmetry~\cite{CDV07} .

\section{Formalism}

In the present work we assume a $\mathcal{PT}$ -symmetric square well
potential

\begin{equation}
V\left( x\right) =\left\{ 
\begin{array}{c}
0\,,\;\hspace{0.5cm}\hspace{0.1cm}\hspace{0.5cm}\hspace{0.5cm}x<-b\;\;(I) \\ 
q\left( V_{0}-iV_{1}\right) \,,\;-b\leq x<0\,\;(II) \\ 
q\left( V_{0}+iV_{1}\right) \,,\;0<x\leq +b\;(III) \\ 
0\,,\;\hspace{0.5cm}\hspace{0.1cm}\hspace{0.5cm}\hspace{0.5cm}x>+b\;\;(IV)%
\end{array}%
\right. \;^{\prime }  \label{PT_well}
\end{equation}%
where the real and imaginary depths, $V_{0}$ and $V_{1}$, respectively, and
the half-width, $b$, are positive numbers , while the elementary charge is
assumed to be $q$ $=-1$ for particles, as the time component of a vector
potential in the Dirac equation in $(1+1)$ dimensions 
\begin{equation}
i\frac{\partial }{\partial t}\Psi (x,t)=H_{D}(q)\Psi (x,t)\,.  \label{Dirac}
\end{equation}

Here, the Dirac Hamiltonian, $H_{D}$, reads 
\begin{equation}
H_{D}=V(x)-i\alpha_{x}\frac{\partial}{\partial x}\;+\beta m.
\label{Dirac_eq}
\end{equation}

Formula (\ref{Dirac_eq}) is written in natural units, $\hbar =c=1$, which
will be used throughout the present work, and the metric is $%
g_{00}=-g_{11}=+1$. The $2\times 2$ $\ $Hermitian matrices $\alpha _{x}$ and 
$\beta $ anticommute and are traceless with square unity: it is thus
possible to identify them with two of the Pauli matrices: the choice we make
corresponds to the standard Dirac representation\cite{MS87}%
\begin{equation}
\alpha _{x}=\sigma _{x}\,,\;\beta =\sigma _{z}\,,  \label{Pauli}
\end{equation}%
particularly suited to the study of the nonrelativistic limit of Eq. (\ref%
{Dirac}).

As is evident from the formulae given above, the solution, $\Psi$, to the
Dirac equation \ref{Dirac} in (1+1) dimensions can be written as a spinor
with two components. The parity operator, $\mathcal{P}$, and the time
reversal operator, $\mathcal{T}$, are to be defined in a consistent way. In
the adopted Dirac representation, we obtain\cite{dC03}%
\begin{equation}
\mathcal{P=\,}e^{i\theta_{\mathcal{P}}}P_{0}\sigma_{z}\;,  \label{P}
\end{equation}
where $\theta_{\mathcal{P}}$ is an arbitrary constant, and $P_{0}$ changes $%
x $ into $-x$. The Pauli matrix $\sigma_{z}$ ensures that the upper and
lower components of $\Psi$ have opposite parities. With formula (\ref{P}) as
definition of the parity operator, it is immediate to check that $\Psi_{%
\mathcal{P}}\left( x,t\right) \equiv\mathcal{P}\Psi\left( x,t\right) $ is a
solution to the Dirac equation (\ref{Dirac}) with potential $\mathcal{P}%
V\left( x\right) \mathcal{P}^{-1}=V\left( -x\right) $.

For the time reversal operator, $\mathcal{T}$, we consistently adopt the
following form 
\begin{equation}
\mathcal{T}=e^{i\theta_{\mathcal{T}}}\sigma_{z}\mathcal{K\;},  \label{T}
\end{equation}
where $\theta_{\mathcal{T}}$ is a constant and $\mathcal{K}$ performs
complex conjugation. $\Psi_{\mathcal{T}}\left( x,t\right) \equiv\mathcal{T}%
\Psi\left( x,t\right) $ satisfies the equation%
\begin{equation}
-i\frac{\partial}{\partial t}\Psi_{\mathcal{T}}\left( x,t\right) =\left(
V^{\ast}\left( x\right) -i\sigma_{x}\frac{\partial}{\partial x}+m\sigma
_{z}\right) \Psi_{\mathcal{T}}\left( x,t\right) \;.  \label{Mod_Dir}
\end{equation}

For simplicity's sake, we may assume $\theta_{\mathcal{T}}=-\theta _{%
\mathcal{P}}$, so that%
\begin{equation}
\mathcal{PT=}P_{0}\mathcal{K\;},  \label{PT}
\end{equation}
since $\sigma_{z}^{2}$ is the identity matrix. Definition (\ref{PT}) is
consistent with the one commonly adopted in nonrelativistic quantum
mechanics (see, \textit{e.g.}, section 4.1 of Ref.\cite{CDV07}).

Moreover, it is easy to show that the operator%
\begin{equation}
\mathcal{C}^{\prime}\mathcal{=}e^{i\theta_{\mathcal{C}^{\prime}}}\sigma_{y}%
\;,  \label{C}
\end{equation}
with $\theta_{\mathcal{C}}$ a real number, meets the condition%
\begin{equation}
\mathcal{C}^{\prime}H_{D}\left( q\right) \mathcal{C}^{\prime-1}=-H_{D}\left(
-q\right) \;.  \label{CHC}
\end{equation}

Therefore, $\Psi_{\mathcal{C}}\left( x,t\right) \equiv\mathcal{C}^{\prime
}\Psi\left( x,t\right) $ fulfills a modified Dirac equation 
\begin{equation}
-i\frac{\partial\Psi_{\mathcal{C}^{\prime}}\left( x,t\right) }{\partial t}%
=H_{D}\left( -q\right) \Psi_{\mathcal{C}^{\prime}}\left( x,t\right) \;.
\label{antiD}
\end{equation}

Note that $\mathcal{C}^{\prime}\mathcal{PT}$ meets condition (\ref{CHC}),
too, provided that the Dirac Hamiltonian, $H_{D}$, commutes with $\mathcal{PT%
}$, \textit{i.e. }$V\left( x\right) =V^{\ast}\left( -x\right) $.

It is worthwhile to point out that the above definitions are different from
those commonly adopted in textbooks\cite{Gr97}. In particular, the
transformed wave function $\Psi_{\mathcal{C}^{\prime}\mathcal{PT}}\left(
x,t\right) \equiv\mathcal{C}^{\prime}\mathcal{PT}\Psi\left( x,t\right) $
satisfies the Dirac equation for "antiparticles"%
\begin{equation}
i\frac{\partial}{\partial t}\Psi_{\mathcal{C}^{\prime}\mathcal{PT}}\left(
x,t\right) =H_{D}\left( -q\right) \Psi_{\mathcal{C}^{\prime}\mathcal{PT}%
}\left( x,t\right) \;.  \label{antipeq}
\end{equation}

In each of the four regions of the $x$ axis defined by formula (\ref{PT_well}%
), we search for particular solutions, $\Phi(x,t)=\Phi_{0}\left( x\right)
e^{-iEt}$, whose spatial part, $\Phi_{0}\left( x\right) $, can be written in
the compact form 
\begin{equation}
\Phi_{0}(x)=u_{\pm}\left( k\right) \cdot e^{\pm ikx}=\left( 
\begin{array}{c}
u_{\pm}^{u}\left( k\right) \\ 
u_{\pm}^{l}\left( k\right)%
\end{array}
\right) \cdot e^{\pm ikx}\,.  \label{Psi}
\end{equation}

Direct replacement of formula (\ref{Psi}) in Eqs. (\ref{Dirac}-\ref{Dirac_eq}%
) yields for momentum $k$ 
\begin{equation}
k^{2}\left( x\right) =\left( E-V(x)\right) ^{2}-m^{2}  \label{K^2}
\end{equation}
and for the ratio, $\lambda$, of lower and upper components%
\begin{equation}
u_{\pm}^{l}=\pm\frac{k(x)}{E-V(x)+m}u_{\pm}^{u}\equiv\pm\lambda(x)u_{%
\pm}^{u}\,,  \label{u_K}
\end{equation}
where the upper components, $u_{\pm}^{u}$, turn out to be arbitrary non-zero
constants, set to 1 for convenience. Adopting the matrix notation of Ref.%
\cite{MS87} , the general stationary solution, $\Psi_{J}\left( x\right) $,
to the Dirac equation in the $J$-th region of the $x$ axis ($J=I,...,IV$)
can be written in the form%
\begin{equation}
\Psi_{J}\left( x\right) =\Omega_{J}\left( x\right) \left( 
\begin{array}{c}
A_{J} \\ 
B_{J}%
\end{array}
\right) \;,  \label{Psi_J}
\end{equation}
where $A_{J}$ and $B_{J}$ are constant and 
\begin{equation}
\Omega_{J}\left( x\right) \equiv\left( 
\begin{array}{cc}
1 & 1 \\ 
\lambda_{J} & -\lambda_{J}%
\end{array}
\right) \cdot\left( 
\begin{array}{cc}
e^{ik_{J}x} & 0 \\ 
0 & e^{-ik_{J}x}%
\end{array}
\right) =\left( 
\begin{array}{cc}
1 & 1 \\ 
\lambda_{J} & -\lambda_{J}%
\end{array}
\right) \cdot e^{ik_{J}\sigma_{z}}\;,  \label{Omega_J}
\end{equation}
is the matrix whose columns are the two linearly independent solutions $%
\Phi_{J}(x)$ in region $J$, apart from a normalization factor, \ which does
not affect the derivation of transmission and reflection coefficients. Here, 
$\lambda_{J}\equiv k_{J}/(E-V_{J}+m)$ and $k_{J}=\sqrt{(E-V_{J})^{2}-m^{2}}$%
. It is worthwhile to notice that $\lambda_{I}=\lambda_{IV}\equiv\lambda
=k/\left( E+m\right) $, with $k=\sqrt{E^{2}-m^{2}}>0$, for scattering
states, since $V_{I}=V_{IV}=0$, while $\lambda_{II}=\lambda_{III}^{\ast}%
\equiv\Lambda$ and $k_{II}=k_{III}^{\ast}=K$, since $V_{II}=V_{III}^{\ast}$.

By imposing continuity of the general solution, $\Psi$, at the boundary, $%
x_{b}^{J}$, between regions $J$ and $J+1$%
\begin{equation*}
\begin{array}{c}
\Omega_{J+1}\left( x_{b}^{J}\right) \cdot\left( 
\begin{array}{c}
A_{J+1} \\ 
B_{J+1}%
\end{array}
\right) =\Omega_{J}\left( x_{b}^{J}\right) \cdot\left( 
\begin{array}{c}
A_{J} \\ 
B_{J}%
\end{array}
\right) \\ 
\Rightarrow\left( 
\begin{array}{c}
A_{J+1} \\ 
B_{J+1}%
\end{array}
\right) =\Omega_{J+1}^{-1}\left( x_{b}^{J}\right) \cdot\Omega_{J}\left(
x_{b}^{J}\right) \cdot\left( 
\begin{array}{c}
A_{J} \\ 
B_{J}%
\end{array}
\right) \;,%
\end{array}%
\end{equation*}
it is easy to express the coefficients of the general solution in region $IV$
($x\rightarrow+\infty$) \ as linear functions of those in region $I$ ($%
x\rightarrow-\infty$)%
\begin{equation}
\left( 
\begin{array}{c}
A_{IV} \\ 
B_{IV}%
\end{array}
\right) =M^{D}\cdot\left( 
\begin{array}{c}
A_{I} \\ 
B_{I}%
\end{array}
\right) \;,  \label{IVtoI}
\end{equation}

where%
\begin{equation}
M^{D}=\Omega_{IV}^{-1}\left( +b\right) \cdot\Omega_{III}\left( +b\right)
\cdot\Omega_{III}^{-1}\left( 0\right) \cdot\Omega_{II}\left( 0\right)
\cdot\Omega_{II}^{-1}\left( -b\right) \cdot\Omega_{I}\left( -b\right)
\label{M_D}
\end{equation}
is the Dirac matching matrix\cite{MS87}. Since $\det\Omega_{J}=-2\lambda_{J}$
and $\det\Omega_{J}^{-1}=1/\det\Omega_{J}$ are independent of $x$, it is
immediate to check that%
\begin{equation}
\det M^{D}=\det\Omega_{I}(-b)/\det\Omega_{IV}\left( +b\right) =\lambda
_{I}/\lambda_{IV}=1\;.  \label{det_M_D}
\end{equation}

It is worthwhile to mention that, in the matching matrix method applied to
the $\mathcal{PT}$-symmetric square well in the one-dimensional Schr\"{o}%
dinger equation in a previous work of ours\cite{CDV07} , use was made of a
matching matrix, $M$, which expresses the coefficients of the general
solution in region $I$ in terms of those in region $IV$; therefore, the $M$
matrix of Ref.\cite{CDV07} is the nonrelativistic limit of $\left(
M^{D}\right) ^{-1}$, easily derivable from formula (\ref{M_D}).

Elementary quantum mechanics immediately yields transmission and reflection
coefficients in terms of $M^{D}$ matrix elements. For a plane wave
travelling from left to right $\left( L\rightarrow R\right) $, we must have $%
B_{IV}=0$, so that, from formulae (\ref{IVtoI}-\ref{M_D})%
\begin{equation}
T_{L\rightarrow R}=\frac{A_{IV}}{A_{I}}=\frac{\det M^{D}}{M_{22}^{D}}=\frac {%
1}{M_{22}^{D}}\;,  \label{T_LR}
\end{equation}
and%
\begin{equation}
R_{L\rightarrow R}=\frac{B_{I}}{A_{I}}=-\frac{M_{21}^{D}}{M_{22}^{D}}\;.
\label{R_LR}
\end{equation}

For a plane wave travelling from right to left $\left( R\rightarrow L\right) 
$, $A_{I}=0$, so that%
\begin{equation}
T_{R\rightarrow L}=\frac{B_{I}}{B_{IV}}=\frac{1}{M_{22}^{D}}\;,  \label{T_RL}
\end{equation}
and 
\begin{equation}
R_{R\rightarrow L}=\frac{A_{IV}}{B_{IV}}=\frac{M_{12}^{D}}{M_{22}^{D}}\;.
\label{R_RL}
\end{equation}
Thus, for a $\mathcal{PT}$-symmetric square well, $T_{L\rightarrow
R}=T_{R\rightarrow L}$, while $R_{L\rightarrow R}\neq R_{R\rightarrow L}$.
The nonrelativistic case\cite{CDV07} suggests the equality of the two
transmission coefficients as a consequence of the intertwining relation $%
\mathcal{T}H=H^{\dag}\mathcal{T}$, satisfied by the Hamiltonian (\ref%
{Dirac_eq}) with any local $\mathcal{PT}$-symmetric potential. The
intertwining relation would be broken by a non-local potential\cite{CDV07}%
\cite{CV06}; in that case, we would have $T_{L\rightarrow R}\neq
T_{R\rightarrow L}$, too.

Transmission and reflection coefficients are entries of the scattering
matrix, $S$\cite{MS87},\cite{CDV07}%
\begin{equation}
S=\left( 
\begin{array}{cc}
T_{L\rightarrow R} & R_{R\rightarrow L} \\ 
R_{L\rightarrow R} & T_{R\rightarrow L}%
\end{array}
\right) \;.  \label{S_mat}
\end{equation}

As a consequence of formulae(\ref{det_M_D},\ref{T_LR}-\ref{R_RL}), $S$ is
not unitary. As already explained in Ref.\cite{CDV07}, our approach uses $%
\mathcal{PT}$-symmetric potentials as effective potentials, and our purpose
is different from that of Ref.\cite{Jo07}, which considers them as
fundamental and thus searches for a different Hilbert-space metric that
permits conservation of probability. As shown in the same reference,
however, one should face, in this latter approach to scattering by localized
potentials, conceptual problems connected with the non-locality of the
metric that ensures unitarity of the $S$ matrix.

From the definition of transmission and reflection coefficients in terms of $%
M^{D}$ matrix elements, it is easy to check that the determinant meets the
condition $\left\vert \det S\right\vert =1$.

Formulae (\ref{Omega_J}-\ref{M_D}) allow us to obtain compact expressions of
the $M^{D}$ matrix elements and, consequently, of the transmission and
reflection coefficients (\ref{T_LR}-\ref{R_RL}) in terms of real $k$ and $%
\lambda$ defined in region $I$ and complex $K$ and $\Lambda$ defined in
region $II$. For the latter quantities it is convenient to use the
parameterization%
\begin{equation}
K=\sqrt{\alpha_{+}\alpha_{-}}e^{i\left( \varphi_{+}+\varphi_{-}\right)
/2},\;\Lambda=\sqrt{\alpha_{+}/\alpha_{-}}e^{i\left( \varphi_{+}-\varphi
_{-}\right) /2}\;,  \label{K_Lambda}
\end{equation}

with%
\begin{equation}
\alpha_{\pm}=\sqrt{\left( E-qV_{0}\pm m\right) ^{2}+q^{2}V_{1}^{2}}%
,\;\varphi_{\pm}=\arctan\left( qV_{1}/\left( E-qV_{0}\pm m\right) \right) \;.
\label{alpha_phi}
\end{equation}

After some algebra, we obtain%
\begin{equation}
\begin{array}{c}
M_{11}^{D}=e^{-2ikb}\left\{ \left( \frac{\operatorname{Im}\Lambda}{|\Lambda|}\right)
^{2}\cosh\left( 2b\operatorname{Im}K\right) +\left( \frac{\operatorname{Re}\Lambda}{|\Lambda|%
}\right) ^{2}\cos\left( 2b\operatorname{Re}K\right) \right. \\ 
\left. +i\left[ \operatorname{Im}\Lambda\sinh\left( 2b\operatorname{Im}K\right) \left( \frac{%
\lambda^{2}-|\Lambda|^{2}}{2\lambda|\Lambda|^{2}}\right) +\operatorname{Re}%
\Lambda\sin\left( 2b\operatorname{Re}K\right) \left( \frac{\lambda^{2}+|\Lambda|^{2}%
}{2\lambda|\Lambda|^{2}}\right) \right] \right\} \;,%
\end{array}
\label{M_11}
\end{equation}

The $M_{22}^{D}$ matrix element turns out to be the complex conjugate of \ $%
M_{11}^{D}$

\begin{equation}
M_{22}^{D}=\left( M_{11}^{D}\right) ^{\ast}\;,  \label{M_22}
\end{equation}
and the off-diagonal elements read%
\begin{equation}
\begin{array}{c}
M_{12}^{D}=i\left\{ \frac{\operatorname{Re}\Lambda\operatorname{Im}\Lambda }{|\Lambda|^{2}}%
\left[ \cos\left( 2b\operatorname{Re}K\right) -\cosh\left( 2b\operatorname{Im}K\right) %
\right] \right. \\ 
\left. -\operatorname{Re}\Lambda\sin\left( 2b\operatorname{Re}K\right) \left( \frac{%
\lambda^{2}-|\Lambda|^{2}}{2\lambda|\Lambda|^{2}}\right) -\operatorname{Im}%
\Lambda\sinh\left( 2b\operatorname{Im}K\right) \left( \frac{\lambda^{2}+|\Lambda|^{2}%
}{2\lambda|\Lambda|^{2}}\right) \right\}%
\end{array}
\label{M_12}
\end{equation}
and%
\begin{equation}
\begin{array}{c}
M_{21}^{D}=i\left\{ \frac{\operatorname{Re}\Lambda\operatorname{Im}\Lambda }{|\Lambda|^{2}}%
\left[ \cos\left( 2b\operatorname{Re}K\right) -\cosh\left( 2b\operatorname{Im}K\right) %
\right] \right. \\ 
\left. +\operatorname{Re}\Lambda\sin\left( 2b\operatorname{Re}K\right) \left( \frac{%
\lambda^{2}-|\Lambda|^{2}}{2\lambda|\Lambda|^{2}}\right) +\operatorname{Im}%
\Lambda\sinh\left( 2b\operatorname{Im}K\right) \left( \frac{\lambda^{2}+|\Lambda|^{2}%
}{2\lambda|\Lambda|^{2}}\right) \right\} \;.%
\end{array}
\label{M_21}
\end{equation}

In the $V_{1}\rightarrow0$ limit, corresponding to a real square well, the
diagonal matrix elements (\ref{M_11}-\ref{M_22}) reduce to the corresponding
ones of Ref.\cite{MS87}, the off-diagonal elements (\ref{M_12}-\ref{M_21})
differ from those of Ref.\cite{MS87} by phase factors due to the different
choice of the origin of the $x$ axis (left edge of the well in Ref.\cite%
{MS87}, centre of the well in the present work): more precisely, $%
M_{12}^{D}=e^{2ikb}\mathcal{M}_{12}^{D}$\cite{MS87}, $M_{21}^{D}=e^{-2ikb}%
\mathcal{M}_{21}^{D}$\cite{MS87}, as expected\cite{CDV07}. The square moduli
of reflection coefficients are obviously not affected by these phase
differences.

It is also of some interest to compute the nonrelativistic limits of the $%
M^{D}$ matrix elements, in order to compare them with the corresponding
expressions for the $\mathcal{PT}$-symmetric square well in the Schr\"{o}%
dinger equation obtained in Ref.\cite{CDV07}. To this aim, we need the
limits of the basic quantities $k_{J}^{2}$ and $\lambda_{J}^{2}$ for $%
E\rightarrow m+\epsilon$, where $\epsilon$ ($<<m$) is the kinetic energy. In
that limit, $k_{I}^{2}=k_{IV}^{2}=k^{2}\rightarrow2m\epsilon$, $\lambda
_{I}^{2}=\lambda_{IV}^{2}=\lambda^{2}\rightarrow\epsilon/(2m)$, $%
k_{II}^{2}=\left( k_{III}^{2}\right) ^{\ast}=K^{2}\rightarrow2m(\epsilon
-qV_{0}+iqV_{1})$, $\lambda_{II}^{2}=\left( \lambda_{III}^{2}\right) ^{\ast
}=(\epsilon-qV_{0}+iqV_{1})/(2m)$. Using units $2m=1$, as in Ref.\cite{CDV07}%
, $k_{J}^{2}$ and $\lambda_{J}^{2}$ coincide. In the same units, one easily
verifies, after some algebra, that $M^{D}\rightarrow M^{-1} $ of Ref.\cite%
{CDV07}, as expected.

The formulation gives above is suited to the description of scattering
states, with $E<-m$ or $E>+m$. \ The energies of discrete bound states, in
the $-m<E<+m$ range, appear as poles of the transmission coefficient, $%
T_{L\rightarrow R}$ $=T_{R\rightarrow L}$, or, equivalently, as zeros of the 
$M_{22}^{D}$ matrix element. For bound states, $k$ and $\lambda$ in the
asymptotic regions become imaginary, $k=ik^{\prime}\equiv i\sqrt{m^{2}-E^{2}}
$ and $\lambda=i\lambda^{\prime}\equiv ik^{\prime}/(m+E)$. The equation
satisfied by real bound-state energies thus reads%
\begin{equation}
\begin{array}{c}
e^{-2bk^{\prime}\left( E\right) }\left\{ \left( \frac{\operatorname{Im}\Lambda(E)}{%
|\Lambda(E)|}\right) ^{2}\cosh\left( 2b\operatorname{Im}K(E)\right) +\left( \frac{%
\operatorname{Re}\Lambda(E)}{|\Lambda(E)|}\right) ^{2}\cos\left( 2b\operatorname{Re}%
K(E)\right) \right. \\ 
+\left. \operatorname{Im}\Lambda(E)\sinh\left( 2b\operatorname{Im}K(E)\right) \left( \frac{%
\lambda^{\prime2}(E)+|\Lambda(E)|^{2}}{2\lambda^{\prime}(E)|\Lambda(E)|^{2}}%
\right) \right. \\ 
+\left. \operatorname{Re}\Lambda(E)\sin\left( 2b\operatorname{Re}K(E)\right) \left( \frac{%
\lambda^{\prime2}(E)-|\Lambda(E)|^{2}}{2\lambda^{\prime}(E)|\Lambda(E)|^{2}}%
\right) \right\} =0%
\end{array}
\label{BS_eq}
\end{equation}

In the $V_{1}\rightarrow 0$ limit, corresponding to the real square well,
formula (\ref{BS_eq}) reduces to the well-known text-book expression\cite%
{Gr97}.

The scatttering eigenfunctions of the $\mathcal{PT}$-invariant Hamiltonian (%
\ref{Dirac_eq}) are not eigenstates of $\mathcal{PT}$: this corresponds to a
spontaneous breakdown of $\mathcal{PT}$ symmetry. A different scenario would
be obtained if the $\mathcal{PT}$-symmetric square well potential were the
space component of a vector potential: in that case, if one assumes minimal
coupling, $p_{x}\longrightarrow p_{x}+qV_{x}$, the time-independent Dirac
equation reads, with our choice of Dirac matrices%
\begin{equation}
E\Psi \left( x\right) =H_{D}^{\prime }\Psi \left( x\right) \equiv \left[
\left( -i\frac{\partial }{\partial x}+qV_{x}(x)\right) \sigma _{x}+\sigma
_{z}m\right] \Psi \left( x\right) \;,  \label{Dirac2}
\end{equation}

By assuming for $qV_{x}(x)$ a square well potential of type (\ref{PT_well}),
and repeating the calculations of the $M$ matrix as in the previous case,
one would obtain the following relation between the coefficients of the
asymptotic solutions in region $I$ ($x\rightarrow -\infty $) and $IV$ ($%
x\rightarrow +\infty $)%
\begin{equation}
\left( 
\begin{array}{c}
A_{IV} \\ 
B_{IV}%
\end{array}%
\right) =e^{-2iqV_{0}b}\left( 
\begin{array}{c}
A_{I} \\ 
B_{I}%
\end{array}%
\right) \;.  \label{Dir_mat_2}
\end{equation}

For a wave travelling from left to right ($L\rightarrow R$), with the
boundary conditions $A_{I}=1$, $B_{IV}=0$, formula (\ref{Dir_mat_2}) yields $%
T_{L\rightarrow R}=A_{IV}=e^{-2iqV_{0}b}$, $R_{L\rightarrow
R}=B_{I}=0$, while, for a wave travelling from right to left ($%
R\rightarrow L$), with $B_{IV}=1$, $A_{I}=0$, one gets $%
T_{R\rightarrow L}=B_{I}=e^{2iqV_{0}b}$, $R_{R\rightarrow
L}=A_{IV}=0 $. The square well potential thus becomes reflectionless and
conserves the probability flux. A more general $\mathcal{PT}$-symmetric
local potential of finite range, $V(x)=V^{\ast }\left( -x\right) $, $\left(
-b\leq x+b\leq +b\right) $, would maintain this property, since, in that
case, the argument $2qV_{0}b$ of the exponential in formula (\ref{Dir_mat_2}%
) would be replaced with the real integral $q\int_{-b}^{+b}V_{R}(x)dx$,
where $V_{R}(x)$ is the real part of $V$, owing to the fact that the
imaginary part is an odd function of $x$ and does not contribute to it.

It is an easy matter to check that the asymptotic wave functions are
eigenstates of $\mathcal{PT}$, in keeping with the proof given in Ref.\cite%
{CDV07} that an exact asymptotic $\mathcal{PT}$ symmetry necessarily implies
that the potential is reflectionless and conserves unitarity. It is
worthwhile to note that in this case $T_{R\rightarrow L}\neq
T_{L\rightarrow R}$, as a consequence of the fact that $\mathcal{T}%
H_{D}^{\prime }\neq H_{D}^{\prime \dag }\mathcal{T}$\cite{CDV07}. The
connection between exact asymptotic $\mathcal{PT}$ symmetry \ and potential
reflectionlessness was proved in Ref.\cite{CDV07} for finite range
potentials, but it might be more general, since it was already pointed out
in Ref. \cite{ABB05} for infinite range potentials of the class $V\left(
x\right) =-x^{2K+2}$ $\left( K=1,2,3,...\right) $.

\section{Results and Comments}

In order to explore scattering properties of an overcritical $\mathcal{PT}$%
-symmetric square well, we have performed several calculations of
transmission and reflection coefficients as functions of an increasing
imaginary depth, $V_{1}$, while keeping real depth, $V_{0}>2m$, and
half-width, $b$, fixed. In our calculations, $V_{0}$ and $V_{1}$ are
expressed in unit of particle mass, $m$, and $b$ in Compton wavelengths $%
\lambda _{C}=1/m$.
\begin{figure}[h]
\begin{center}
\includegraphics[width=10cm,angle=270,clip]{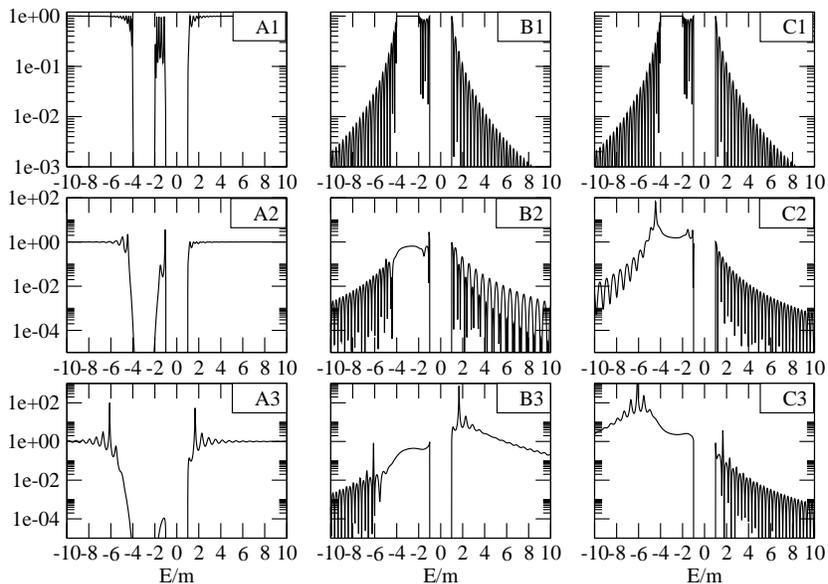}
\caption{Transmission and reflection coefficients for $\mathcal{PT}$-symmetric
square wells with $q=-1$,$b=5/m$,$V_0=3m$. 1st row: $V_1=0$; 2nd row: $V_1=0.25m$;
 3rd row: $V_1=0.5m$. A1-3 : $|T|^{2}$; B1-3: $|R_{L\rightarrow R}|^{2}$;
 C1-3: $|R_{R\rightarrow L}|^{2}$.} 
\label{Fig1}
\end{center}
\end{figure}
As an example of our results, Figure 1 gives $|T_{L\rightarrow R}|^{2}$ (= $%
|T_{R\rightarrow L}|^{2}\equiv|T|^{2}$), $|R_{L\rightarrow R}|^{2}$ and $%
|R_{R\rightarrow L}|^{2}$ versus energy, $E$, with $m=1$, $V_{0}=3$, $b=5$
and $V_{1}=0$ (top panels), $V_{1}=0.25$ (intermediate panels) and $%
V_{1}=0.5 $ (bottom panels). As a general comment, the signature of
spontaneous pair production inside the well is represented by the
transmission resonances at negative energies in the $-2\leq E/m\leq-1$
range, adjacent to the bound-state region ($-1<E/m<+1$), where the
transmission coefficient may have poles on the real axis, corresponding to
bound states. The half-plane $E/m>+1$ corresponds to positive-energy
scattering, the half-plane $E/m<-4$ ($=-1-V_{0}/m$) to negative-energy
scattering.

The top panels refer to a real well $(V_{1}=0)$. In this case, the two
reflection coefficients are equal : $|R_{L\rightarrow
R}|^{2}=|R_{R\rightarrow L}|^{2}\equiv |R|^{2}$ and unitarity holds: $%
|T|^{2}+|R|^{2}=1$. The intermediate panels refer to a $\mathcal{PT}$%
-symmetric well with $V_{1}=0.25m$ : spontaneous pair creation is still
sizable, the two reflection coefficients differ by more than an order of
magnitude at their maxima and unitarity is broken. The bottom panels refer
to a $\mathcal{PT}$-symmetric well with $V_{1}=0.5m$: with increasing $V_{1}$%
, spontaneous pair creation is drastically reduced, albeit still present;
the two reflection coefficients differ by order of magnitudes and $%
|R_{L\rightarrow R}|^{2}$ is sharply peaked at positive energies (particle
reflection), $|R_{R\rightarrow L}|^{2}$ at negative energies (antiparticle
reflection). The potential is neither absorptive ($\left\vert
T_{i\rightarrow j}\right\vert ^{2}+\left\vert R_{i\rightarrow j}\right\vert
^{2}<1$ at all incident energies), nor generative ($\left\vert
T_{i\rightarrow j}\right\vert ^{2}+\left\vert R_{i\rightarrow j}\right\vert
^{2}>1$ at all energies), but shows an intermediate behaviour.

As for bound states, the real well considered in our example has four bound
states at negative energies and four at positive energies, which still
persist when the imaginary part is weak, such as $V_{1}=0.25m$, considered
in the intermediate panels of Fig.1. Real bound states begin to disappear at
a critical value $V_{1crit}\simeq0.272m$. At $V_{1}=0.5m$ bound states with
real energies do not exist any more. The behaviour of the real bound state
spectrum with increasing $V_{1}$ appears to be consistent with a spontaneous
breakdown of $\mathcal{PT}$ symmetry, which deserves, in any case, a more
detailed formal treatment in relativistic quantum mechanics, including a
discussion of spectral degeneracies.

The present work should be considered as a phenomenological exploration of
properties of $\mathcal{PT}$-symmetric local vector potentials in standard
relativistic quantum mechanics; as such, it might be easily extended to
non-local potentials\cite{CV07}. Further developments could involve
either the study of the $\mathcal{C}$ operator~\cite{BT06} 
when $\mathcal{PT}$ symmetry holds , or the
search for a more general symmetry than $\mathcal{PT}$ \ for a Dirac
equation that is not $\mathcal{PT}$-symmetric, as discussed, for instance,
in Ref.\cite{CCZV05}, for the Schr\"{o}dinger equation.

\end{document}